\def\mytitle{Article Title: Smart Education \vspace{20pt} }

\def\mysubtitle{Higher Education Instruction and the Internet of Things (IoT)}

\def\mykeywords{internet of things, education, connectivity, smart worlds, smart classrooms, learning, digital era}

\def\myauthor{ Idris Skloul Ibrahim and Benjamin Kenwright }

\documentclass[11pt]{article}

\usepackage{natbib}
\let\cite\citep

\usepackage{multicol}

\usepackage[backref=page]{hyperref}

\hypersetup{
    colorlinks = true, 
    linkcolor = blue,
    anchorcolor = red,
    citecolor = blue, 
    filecolor = red, 
}

\hypersetup{
    bookmarks=true,         
    unicode=false,          
    pdftoolbar=true,        
    pdfmenubar=true,        
    pdffitwindow=false,     
    pdftitle={\mytitle},    
    pdfauthor={\myauthor},     
    pdfsubject={\mytitle},   
    pdfcreator={\myauthor},   
    pdfproducer={},  
    pdfkeywords 		= {\mykeywords}, 
    pdfnewwindow=true,      
    colorlinks=false,       
    linkcolor=red,          
    citecolor=green,        
    filecolor=magenta,      
    urlcolor=cyan           
}

\usepackage{bookmark}

\usepackage{multicol}

\usepackage{url}

\usepackage{atbegshi,picture}

\title{\mytitle \\ \mysubtitle }
\author{\myauthor } 

\usepackage{enumitem}
\setitemize{noitemsep,topsep=0pt,parsep=0pt,partopsep=0pt,leftmargin=*}
\setenumerate{noitemsep,topsep=0pt,parsep=0pt,partopsep=0pt,leftmargin=*}

\usepackage[compact]{titlesec}

\usepackage{enumitem}
\setitemize{noitemsep,topsep=0pt,parsep=0pt,partopsep=0pt,leftmargin=*}
\setenumerate{noitemsep,topsep=0pt,parsep=0pt,partopsep=0pt,leftmargin=*}

\usepackage{tikz}
\usepackage{calc}
\usepackage{setspace}

\usetikzlibrary{mindmap,shadows}

\usepackage{pgfplots}

\usepackage{pgf-pie}

\usepackage{smartdiagram}

\usepackage{bchart}

\usepackage{graphicx}

\graphicspath{{./images/}}

\newcommand{\figuremacroW}[4]{
	\begin{figure}[!h] 
		\centering
		\includegraphics[width=#4\columnwidth]{#1}
		\caption[#2]{\textbf{#2} - #3}
		\label{fig:#1}
	\end{figure}
}

\usepackage{pifont} 

\usepackage{wasysym} 

\usepackage{amssymb}

\def\mytickbox2{$\text{\rlap{$\checkmark$}}\square$}
\def\mytickbox{$\text{\rlap{\ding{52}}}\square$}

\begin{document}

\widowpenalties 1 1
\raggedbottom
\sloppy

\date{}

\maketitle

\vspace{-20pt}


\section*{\centering\Large Abstract}

The Internet of Things (IoT) has many applications in our daily lives. One aspect in particular is how the IoT is making a substantial impact on education and learning; as we move into the `Smart Educational' era.
This article explores how the IoT continues to transform the education landscape, from classrooms and assessments to culture and attitudes.
Smart Education is a pivotal tool in the fight to meet the educational challenges of tomorrow. %
The IoT tools are getting used more and more often in the area of education, aiming to increase student engagement, satisfaction and quality of learning.
IoT will reshape student culture and habits beyond belief.
As Smart Education is more than just using technologies, it involves a whole range of factors, from the educational management through to the pedagogical techniques and effectiveness. %
Educators in the 21st century now have access to gamification, smart devices, data management, and immersive technologies. 
Enabling academics to gather a variety of information from students.
Ranging from monitoring student engagement to adapting the learning strategies for improved learning effectiveness. %
Through Smart Education, educators will be able to better monitor the needs of individual students and adjust their learning load correspondingly (i.e., optimimal learning environment/workload to support and prevent students failing).
One of the biggest challenges for educators is how new technologies will address growing problems (engagement and achievement).
The scale and pace of change (technological IoT era) is unprecedented.
Typically, jobs students are trained for today will not be here tomorrow.
Education is not just about knowledge acquisition, but also the digital skills, adaptability and creativity (essential, if students are to thrive in the new world).

\vspace{10pt}

\textbf{Keywords: \mykeywords}
\\


\newpage

\begin{figure*}
\centering
\scalebox{0.7}
{
\begin{tikzpicture}[grow cyclic, text width=2cm, align=flush center, every node/.style=concept, concept color=orange!40,
level 1/.style={level distance=4cm,sibling angle=60},
level 2/.style={level distance=4cm,sibling angle=45}]

\node{Internet of Things (IoT)}
   child [concept color=blue!30] { node {Smart Cities }
        child { node {Smart Homes} }
   }
    child [concept color=teal!40]  { node { Smart Roads }
        child { node {Smart Cars}}
    }
    child [concept color=purple!10] { node {Smart Wearable}
        child { node {Smart Watches}}
        child { node {Smart Clothes}}
        child { node {Smart Glasses}}
        child { node {Smart Hearing Aids}}
    }
    child [concept color=yellow!30] { node { Smart Agriculture }
        child { node {Smart Farming}}
    }
    child [concept color=red!50] { node [concept,circular glow={fill=orange!50},scale=1.2] {\textbf{Smart Education}}
        child { node {Smart Classrooms}}
        child { node {Smart Campuses}}
    }
    child [concept color=green!20] { node {Smart Hospitals}
        child { node {Smart Instruments}}
        child { node {Smart Doctors (Avatars)}}
    };

\end{tikzpicture}
}
\caption{Smart IoT World - As the integration and connection of devices continues to grow - we head towards an ever more IoT-driven ``smart world''. Smart devices in every aspect of life - from the home to the hospital, not to mention, in education.
}
\end{figure*}
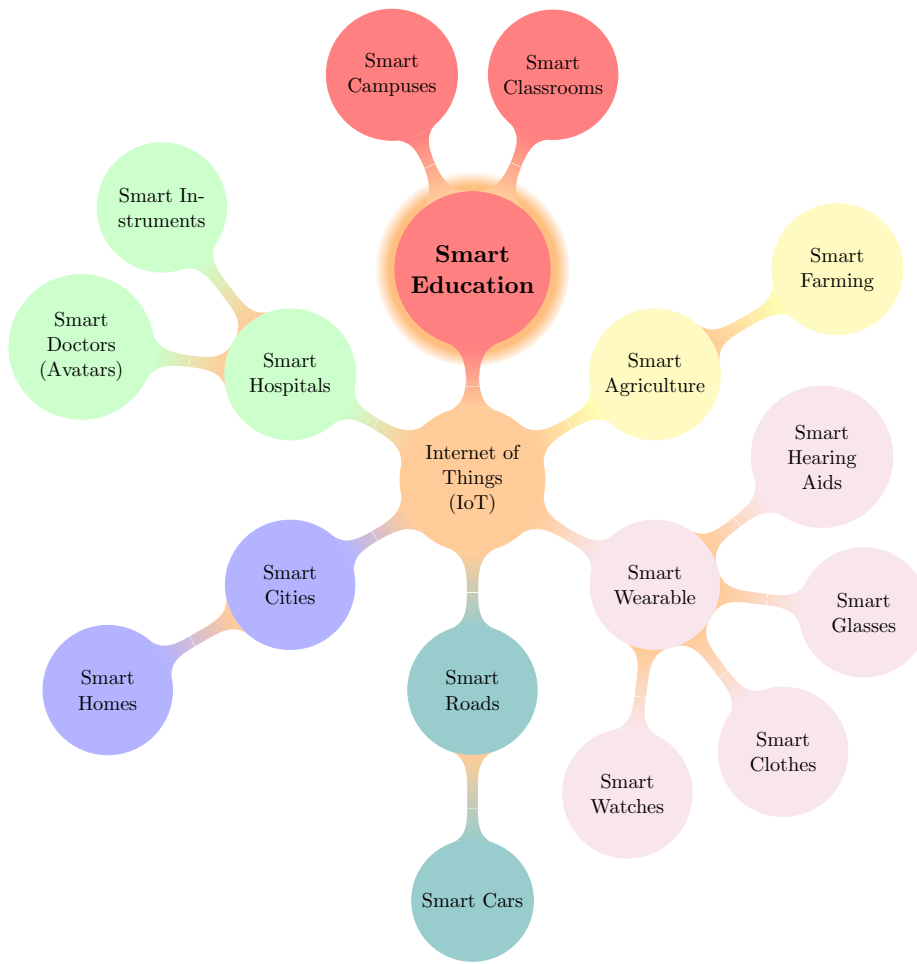

\section{Introduction}
\paragraph{Smart World}

The Internet of Things (IoT) operates within a rapidly changing industry built on innovation and cutting edge technologies \cite{bhayani2016internet,zhu2015green}.
These technological advancements drive smart environments (or `pervasive computing') which permeates many spaces of our daily live.
Smart Hospitals \cite{zhang2018connecting}, Smart Farming \cite{kamilaris2016agri} and Smart Cities \cite{zanella2014internet} (e.g., like airports, hospitals or university campuses) are already equipped with a mass of connected devices (various types and complexity).
One particular area that has started to take hold in recent years, is the envisioned `Smart University' and `Smart Education' which uses the opportunities provided by pervasive computing technologies to benefit learning (i.e., students and staff) \cite{alvarez2017smart}. 
In such a setting, smart technologies in learning environments are able to address current and future pedagogical challenges e.g. engagement, effective communication complex ideas, cost effectiveness and disabilities. 
`Smartness' centered on people (end users).
A particular concern is how to incorporation the IoT technologies in a learning environment. 
This involves devising new concepts for making software applications and services more aware of their learner's needs (engage and adapt as needed).

Questions this article addresses:
\begin{itemize}
    \item Do we need to change from instructors to coaches?
    \item How does the IoT complement traditional educational values?
    \item What are the benefits to students and staff?
    \item What is the wider social and moral impact of the IoT? (are people more isolated)
    \item What are the hidden dangers and opportunities? (data privacy and security)
\end{itemize}

\paragraph{Contribution}
The key contributions of this article are:
(1) review of the IoT with respect to education;
(2) what new possibilities and associated challenges are on the horizon;
(3) we discuss the impact on traditional educational paradigms;
and
(4) where is the internet of education going (today and tomorrow) based upon trends and patterns in the literature.

\begin{figure}
\input{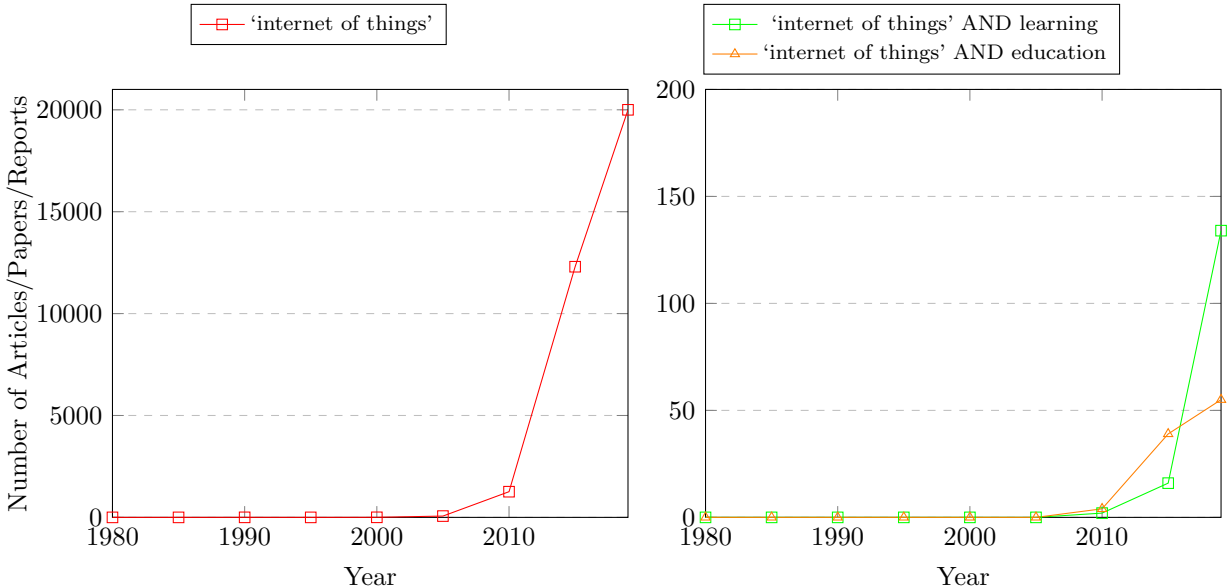}
\caption{A coarse guide to the number of articles in the area of internet of things.  Plots show the number of new articles with the keywords in their title published over the past few decades (from Google Scholar 29/08/2019))].}
\label{fig:plotpublications}
\end{figure}

\section{Education and Technology}

\paragraph{Why is Education Changing?}
There is a meaningful change in education, as society and the world evolves (moving into a new era).
This influences how people learn and interact with the world around them (and in the classroom).
Influences culture, attitude and values.
Take part in the learning.
Information (digital) age.
New tools and new ways of thinking.
Access to a plathora of information at a click of a button.
The instructor no longer holds all the `knowledge'.
Instructors are coaches (guide students on how they should learn).
Students must learn skills and knowledge that makes them useful for tomorrows world (i.e., work with fast changing technologies).
Shift in how people learn \cite{hanna2000higher}.
The world has and continues to change, as we enter into the fourth industrial evolution and the rise of the digital era, we face increasing economic uncertainty with global political and social instabilities.

\paragraph{Well-being and Comfort}
Education is more successful when students are emotionally, socially and spiritually happy.
Well-being and high academic achievement go hand-in-hand \cite{el2010health}.
Supporting students is more than just about making the learning material easily available (environment and experience are also important).
A person must see self-worth in themselves to be successful.
A strong sense of self-worth leads to a positive attitude and is reflected in the students learning and engagement.
This self-worth is achieved through multiple factors, such as, belonging, well-being and even physical aspects (healthy eating and exercise).
This is supported at university, through opportunities, providing students with social events, group work, activities and an open safe working environment.

\paragraph{The Student is `King'}

Higher and further educational institutions are under pressure to transform the learning experience.  Student fees and flawed ranking systems, means universities are becoming more competitive.  Research vs teaching focused institutions.  Innovation leading to vocational and flexible opportunities (industry and employability factors).
Explosive growth in educational technologies (social, cloud, mobile and video technologies).
Growing expectations for open interfaces, open data and interoperability.
Increasing use of personal devices and direct access to third-party solutions.

\paragraph{Attendance and IoT}
At the heart of the idea, is if students attend they are more likely to succeed.  However, research indicates that the issues for non-attendance is more complex \cite{rodgers2002encouraging}.
A renewed and innovative approach to supporting students is required.
Simply tracking students or even fining or punishing students who do not attend is not effective.  While these approaches may offer some-short term improvement, this does not resolve the issues of poor engagement and low student achievement.
At the core of the problem, we must ask ourselves what are the issues, such as:

\begin{itemize}
  \item What motivates students to learn?
  \item How is the students learning connected to their lives (and communities)?
  \item How do students learn best?
  \item What health, family, financial, or personal problems at students struggling with (distract them from their academic focus)?
\end{itemize}

No doubt about it, when a student is truly engaged and focused on their studies, they are unstoppable, they go above and beyond in their duties and work.  This influences their peers, family and instructors.
We must ask how technologies, such as, IoT is able to achieve this, to support each individual student's needs (e.g., remote access, flexible learning).

\figuremacroW
{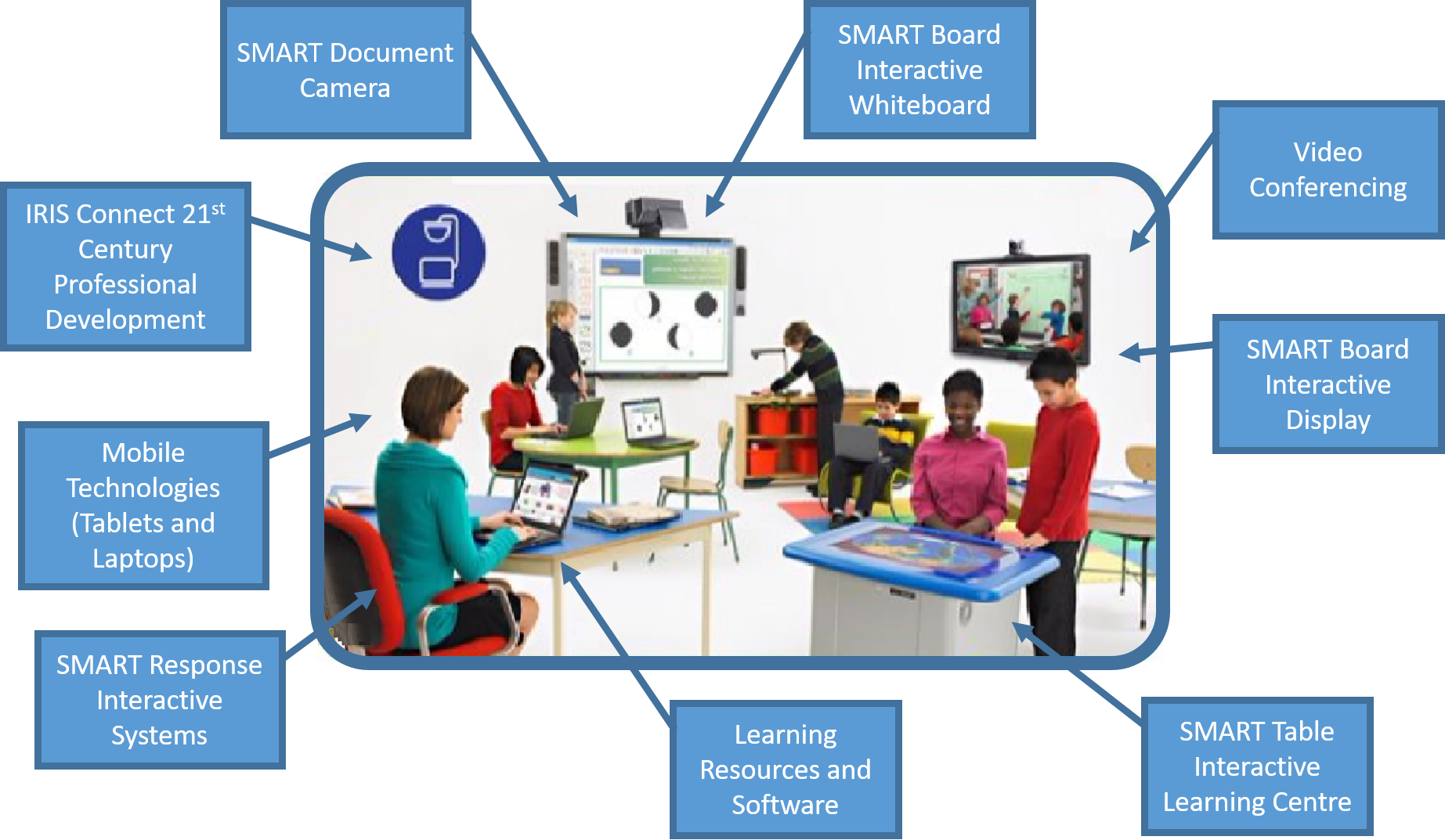}
{Smart Classroom}
{Technologies are getting used more and more in the classroom to facilitate flexible learning.  Engaging the communication of understanding in interactive ways.}
{1.0}

\section{Smart Education}

\paragraph{Unlimited Knowledge (Internet)}
Almost a 100\% of universities across the world have access to the internet.
An almost unlimited wealth of information and knowledge.
It must be argued, the internet is arguably one of the most successful and useful tools mankind has ever created.
The Internets benefits in learning and education are limitless.
Students are able to search for information on any topic they are interested in (or find it hard to talk about).
Connected to social media, help them with informed decisions (advice and support).
Online access means there is a great deal of flexibility and learning choices.
Able to connect people `regardless of the distance'.  For example, Skype has connected more than 300 million people around the world (2.6 million Skype calls a year) \cite{wuttidittachotti2015qoe}.
Meaningful shift influencing every area of society, including education, leading to many internet related advancements (e.g., IoT), which impact polices, public opinion, culture and the economy.
The explosion of IoT devices over the past thirty years, has created technological revolution.
The IoT is a necessity for education, employability, finance and more.
The growing use of terms such as, digital divide, digital literacy and digital inclusion all reflect the growing realization that technologies, such as, the IoT, have become irreducible components of modern life.
Significant impact on every aspect of our lives, including educational sector.

\begin{itemize}
\item  Instructor is: a person who teaches a subject or skill: someone who instructs people

\item Coach is: a person who provides formal, professional coaching to improve the persons effectiveness and performance, and help them achieve their full potential
\end{itemize}

\paragraph{Current University Technologies}

Technologies are an efficient way to manage and engage students at their own level of comfort.  Educational intuitions are already using technologies increasingly (both directly and indirectly to support students). 
For example, a short list of current smart technologies in education include:

\begin{multicols}{2}
\begin{itemize}
  \item Cameras and video
  \item Interactive whiteboards
  \item Tablets and eBooks
  \item Bus tracking (transport)
  \item Student ID cards (RFID)
  \item Airplay and smart televisions
  \item 3D printers
  \item Smart podiums
  \item WiFi everywhere (on campus)
  \item Electric lighting
  \item Attendance tracking
  \item Wireless doorlocks (and booking system)
  \item Temperature sensors
\end{itemize}
\end{multicols}

\figuremacroW
{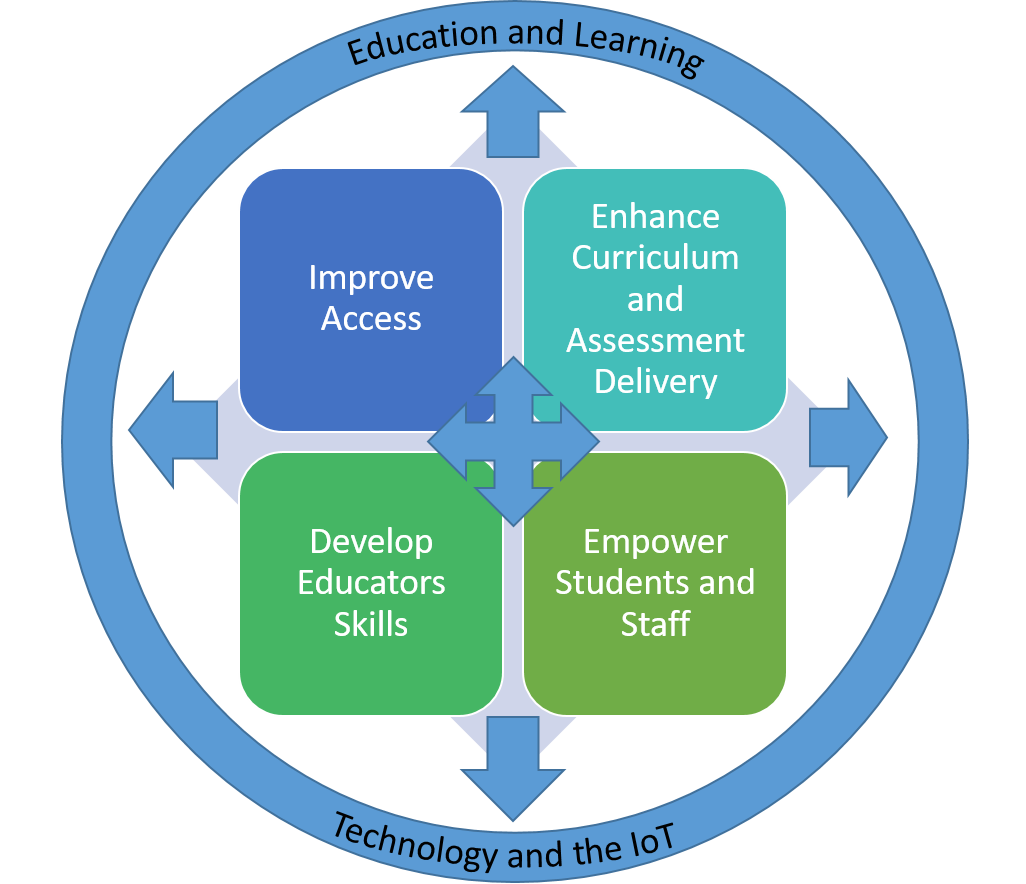}
{Technologies Enhance and Support Education}
{
The IoT impacts education on numerous levels as people move towards a `digital culture'.
Students (and educators) are becoming more confident about using digital technologies.
Leading to new pedagogical approaches and associated tools/infrastructure.
}
{0.6}

\paragraph{Digital Assessment - Barriers and Benefits}
The technologies are not just about communicating and teaching students - the technologies are also impacting how we `assess' students.
Currently, there is a clear move towards digital assessments (e.g. e-exams).
Offers a streamlined system (for delivery and marking).
Not just a matter of converting paper tests to e-versions.
Online assessments and e-assessments are being use more and more in education.
The forms of e-assessment, are anything from multiple choice to drawing a digital diagram.
The assessments can be automatically marked (scripts) or by humans (manually).
The application of digital online assessment are limitless and include a vast range of types (from portfolio to gamification).
What is interesting though, is the technologies and assessments are constantly changing and evolving.
Importantly, a big benefit of online assessment, is it allows greater transparency, security, flexibility and efficiency.
A barrier to digital assessment, is that existing paper exams and approaches cannot merely be transferred over to a digital form.
Peer review is another example, so learners are give the power to critique their peers (build a community and give feedback and help).
Ultimately thought, no one size fits all solution exists for digital assessments, as a biology and art student would have different criteria, with bespoke solutions often being required (avoid hindering or watering down the assessment criteria).

Educational system is changing, no longer simply providing content and assessing that the student has learned and understood the material:

\begin{enumerate}
  \item Student is a Consumer
  \item Students want experience over Knowledge
  \item Education is Changing
  \item Engagement is Interactive
\end{enumerate}

To achieve this, we must embrace new ways of thinking, which include innovative pedagogical practices and immersive and engaging technologies (e.g., IoT, Virtual Reality and Augmented Reality).

\paragraph{Train for Tomorrow}
Many of the jobs we train children and students for today may not exit tomorrow (or the jobs they will do do not exist today). 
The rapid advancement and widespread applications of information technologies provide unpredictable opportunities.
How will we ensure graduates are trained and able to meet the challenges ahead?
The educational sector, needs to ensure students are equiped with the necessary skillsets, which means embracing dynamic strategies for learning and teaching.
The IoT is helping the educational sector meet these challenges head-on (innovations for enhancing the quality of learning which would be difficult for traditional classroom-based approaches).
This opens the door to new innovative pedagogical practices for interactions between things and humans - to enable the realization of smart educational institutions for enhance and improve learning to meet tomorrows jobs.

\figuremacroW
{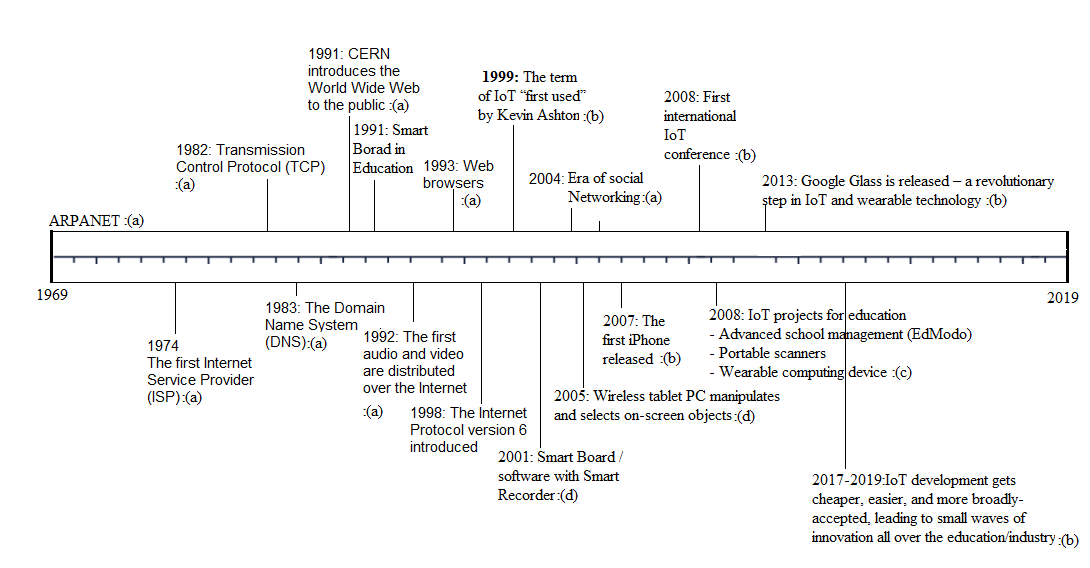}
{IoT time line in Education}
{
Solutions Internet of Things brings in education.
(a) \cite{Internet_History},
(b) \cite{History_of_IoT_A_Timeline_of_Development}.
(c) \cite{IoT_In_Education}.
(d) \cite{History_of_SMART_Board},
}
{1.0}

\section{IoT applications in education sector}
Nowadays, no one able to use and enjoy the benefit of electronic devises and information technology unless connected via a network. We have started to believe that the IoT is playing a part in the rapid change of our society especially in education sector, additionally anyone who denies that isn’t living in the real world. IoT is applied in many sectors and industries (e.g. enterprises, health care, retail, government, teaching, communication) \cite{Huansheng2012,yan2010application,majdub,mse238blog.stanford.edu}.
 
\paragraph{IoT in Teaching and Interactive Learning}
The major reason behind implementing IoT in the education sector, is to enhance the learning environment (e.g. IoT can connect academic sectors all over the world to provide a deeper learning experience for different types of students to gain an easy and rapid high quality education), also to provide advanced value to reconsider how the education components are run. In the next few sections we will touch the most important IoT applications in education sector.

\paragraph{Smart Posters applications} IoT has developed/improved poster boards and are now used as multimedia labels. It is possible to easily create virtual posters that combine images, audio, video, text, and hyperlinks. IoT allows the users (e.g. instructors, students) to share such digital posters in high quality with others and monitor students activities easily. These digital posters can then be shared with classmates and instructors/teachers via email\cite{everythingconnected,majdub}.

\paragraph{Smart board applications} Such applications are able to help teachers to explain lessons more easily in an efficient and interactive techniques with help of using the online presentations and videos. Students in classroom are encouraged to use interactive games as a powerful platform, that because of Web-based applications help to teach students more effectively. Smart technology allows their users such as instructors/teachers and students to browse the web and even edit video clips and sharing the contents interactively (see Figure \ref{fig:smartboard}). 
 
\figuremacroW
{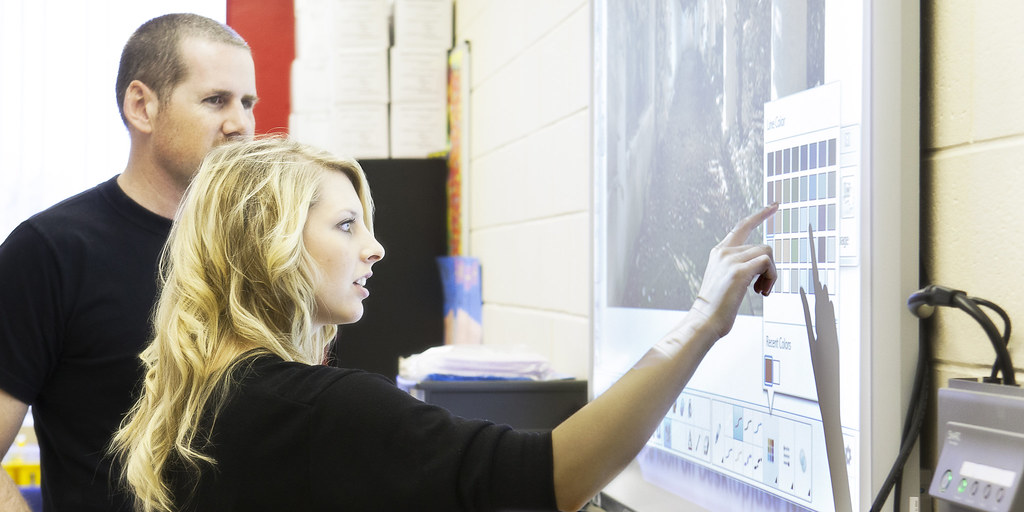}
{Smartboard}
{Improve engagement and interactive experience (vivid displays that are touch sensitive).}
{0.5}
  
\paragraph{Interactive Learning applications} Today's learning is not limited to a combination of text and images but beyond that, most textbooks are available online (loaded to web sites) which include additional materials such as videos, assessments, animations to support effective learning. All that give students a broader view of learning new topics with better understanding and interaction with their teachers and colleagues/partners in classroom and distance learning as well. Additionally, bring real-world problems into the classroom and allow students to find their own solutions. 
 
\paragraph{Sensors and smart devices:} Smart phones, tablets and motion sensors are one of the interesting applications. They are a real change to the teaching and learning field that can be considered a powerful tools which allow students and teachers to create 3D graphics, use e-books that include videos, educational games and give them the ability to take notes, also to provide the best way to learn new topics which makes education more attractive than ever. 
 
\paragraph{Multimedia Digital Library (e-books)}  That provide a better way of learning that allow users (e.g. teachers, students) to carry a library of hundreds of e-books with them easily including graphics, 3D figures, animation and video, smart mobile devices can contain hundreds of textbooks, in addition to homework and other related files and thus eliminate the need for physical storage of books, which contributes to a richer experience and expand learning opportunities for students.  

\paragraph{Sensors in classroom} There are advanced temperature sensors that allow schools to monitor different conditions under any circumstances, which is not only saving thousands in utility costs but also enhancing learning capabilities. Hence, have a significant impact on students 'cognitive skills abilities, memory, attitudes and teachers' feeling. Additionally, using such advanced sensors technology can help to monitor all classrooms remotely  from anywhere.

\section{Privacy and data Protection}
The main challenge of IoT in education is privacy and data protection. Once the data has captured (videos, images) or collected (text, numbers) must be accessible, and secured (protected). Own and accessing a sensitive data is a concerned. Data privacy is an urgent priority in education sector exactly as in finance and defence sectors\cite{abomhara2014security,ValidationDataIntegrity}.

Privacy and data security is the process of protection an education enterprise information and the confidentiality of information from illegal use and threats/intrusions, threats that may threaten the enterprise economically and socially. In fact, maintaining confidentiality and protecting information is only one aspect of security; specialists in digital data security believe that the privacy and data protection consists of the following components \cite{Informationsecurityine-learning}.

\paragraph{Data confidentiality} This includes all necessary measures to prevent unauthorized access to confidentiality (e.g. students’ and staff personal information, academic information, and institution financial status).

\begin{figure}
    \centering
    \input{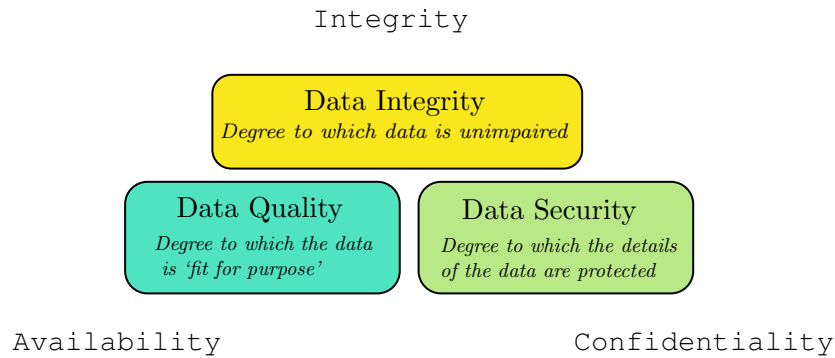}
    \caption{Security vs Integrity (importance of data security in the IoT context).}
    \label{fig:integrity}
\end{figure}

\paragraph{Data integrity} In this aspect it is not important to keep information confidential, but we are concerned about taking measures to protect the information from amending or change (see Figure \ref{fig:integrity}). 
The important question is how do we preserve data integrity?

\paragraph{How to preserve data integrity?}
\begin{itemize}
\item Invalid input: Data can be entered (supplied) by a known or unknown source (end user, an application, a malicious user …etc.). Therefore, implementing checks on errors (e.g. checking human errors when data is entered or transferred from one machine to another) are very important to increase data integrity.
\item Remove duplicate and redundant data: It is a big risk if you use the old file rather than the up to date one. To be in safe side of such risk remove duplicated files.
\item Backup data: It is important to prevent permanent data from lose! Backup is critical when there is a ransomware (malware) attack, backup scheduling best practices to ensure availability.
\item Access controls: Additionally, it is important to minimize the risk of unauthorized access, only the authorized sources can access sensitive data.
\end{itemize}

According to all the above, the risks must be addressed before implement and relay on any technology that may heavily deal with personal or academic student data.

\section{IoT Educational Toolbox: Right Tool for the Right Job}

A number of issues and concerns need to be addressed with respect to the success of Smart Education, such as, is it really better than traditional approaches?
This is, an important area for debate, which is not very often brought up in the literature as an imminent concern, but is important.
Students, instructors and employers need to be concerned with the heterogeneity and complexity of the technologies and the unprecedented impact they have on the educational system and society in general.
A shallow or impaired education system based upon technologies may cause severe harm - both to educational institutions and to the economy.
While normally, in the use of technology for learning, concerns are placed on actions the device take to accomplish their desired task, as specified, education and learning is a concern related to ensuring that the device complements the learning experience (communicate the information in an effective and meaningful way). 
Just like with a regular TV set, which must can be used to communicate educational programmes, but also could be used for entertainment or more, there are multiple instances of IoT applications, however the educational environment could suffer due to
unintended misbehavior of the device.
For examples, internet-enabled devices that may malfunction (and cause consequences), include automated mobile devices that instead of instructing cause confusion to students, a interactive classroom camera causing accidents due to a malfunction of the software, or a smart thermostat in a classroom causing overheating.
There is no doubt that the IoT are a critical part of education and must be taken into account in the design processes to help ensure success. Academia understand the challenges coming with the advent of IoT and are taking steps to improve the design and operational uses. From an instructor perspective the IoT in the context of learning, concerns and proposes establishing guidelines and cases for successful and effective utilization of the IoT. Important issues that we have not discussed in this article, include certification, testing and validation and legal aspects of IoT devices in educational contexts.

\section{Conclusion and Discussion}
In an increasingly connected digital world (the smart generation),
vast opportunities and benefits are available,
transform every aspect of our lives,
especially in education and learning (extraordinary ways).
This smart IoT age, is a catalyst for change, and will impact how we learn, on a global level (influence the economy and society).
Benefit million of people and lead to a wealth of new industries, marketplaces and jobs.
While there are huge opportunities, we must also remember to tread cautiously, for instance, we do not know what the long term deeper emotional and psychological impacts of smart education will be.

\section*{Acknowledgements}
We want to thank the reviewers for taking the time out of their busy schedules to provide insightful and valuable comments to help improve the quality of this article.

\bibliographystyle{apa}

\let\oldthebibliography=\thebibliography
\let\endoldthebibliography=\endthebibliography

\renewenvironment{thebibliography}[1]{%
    \begin{oldthebibliography}{#1}%
      \setlength{\parskip}{0ex}%
      \setlength{\itemsep}{0.5ex}%
  }%
  {%
    \end{oldthebibliography}%
  }

\bibliography{paper}

\end{document}